\documentclass[%
 reprint,
 amsmath,amssymb,
prl,
 longbibliography,
 lengthcheck,%
]{revtex4-1}

\usepackage{graphicx}% Include figure files
\usepackage{dcolumn}% Align table columns on decimal point
\usepackage{bm}% bold math
\usepackage{hyperref}% add hypertext capabilities

\makeatletter

\newcommand{\Rmnum}[1]{\expandafter\@slowromancap\romannumeral #1@}
\makeatother

\begin{document}

\preprint{APS/123-QED}

%CaFe$_2$As$_2$
\title{Magneto-structural coupling and harmonic lattice dynamics in CaFe$_2$As$_2$ probed by M\"ossbauer spectroscopy}% Force line breaks with \\

%\author{Zhiwei Li}
%\author{Xiaoming Ma}
%\author{Xin Liu}
%\author{Tao Wang}\email{lizhiwei03@lzu.cn}
\author{Zhiwei Li}\email{lizhiwei03@lzu.cn}
\author{Xiaoming Ma}
\author{Hua Pang}\email{hpang@lzu.edu.cn}
\author{Fashen Li}%\email{lifs@lzu.edu.cn}
 \affiliation{Institute of Applied Magnetics, Key Lab for Magnetism and Magnetic Materials of the Ministry of Education, Lanzhou University, Lanzhou 730000, Gansu, P.R. China.}%Lines break automatically or can be forced with \\

%\author{Xianhui Chen}
% \affiliation{Hefei National Laboratory for Physical Science at Microscale and Department of Physics, University of Science and Technology of China, Hefei 230026, Anhui, P.R. China. }

\date{\today}% It is always \today, today,

\begin{abstract}
In this paper we present detailed M\"ossbauer spectroscopy study of structural and magnetic properties of the undoped parent compound CaFe$_2$As$_2$ single crystal. By fitting the temperature dependence of the hyperfine magnetic field we show that the magneto-structural phase transition is clearly first-order in nature and we also deduced the compressibility of our sample to be $1.67\times10^{-2}\,GPa^{-1}$. Within the Landau's theory of phase transition, we further argue that the observed phase transition may stem from the strong magneto-structural coupling effect. Temperature dependence of the Lamb-M\"ossbauer factor show that the paramagnetic phase and the antiferromagnetic phase exhibit similar lattice dynamics in high frequency modes with very close Debye temperatures, $\Theta_D \sim$270\,K.
\begin{description}
  %  \item[Usage]
  % Secondary publications and information retrieval purposes.
\item[PACS numbers]
76.80.+y, 74.10.+v
  %  \item[Structure]
  %  You may use the \texttt{description} environment to structure your abstract;
  %  use the optional argument of the \verb+\item+ command to give the category of each item.
\end{description}
\end{abstract}

\maketitle

%\tableofcontents
\section{\label{sec:Intro}Introduction}

The discovery of superconductivity (SC) with critical temperature ($T_c$) up to 55\,K in iron-arsenide systems \cite{26KLaFeAsOF,ReviewNature,ReviewMagn} has triggered enormous interest in iron-based superconducting materials. Generally, the parent compound undergoes a structural transition and exhibits antiferromagnetic (AFM) order below room temperature. Doping with electrons or holes into the parent compound will suppress the structural and magnetic transitions and induce SC \cite{ThePuzzles}. Nowadays it is widely believed that SC is directly coupled to the magnetism in iron-based superconductors \cite{SpinFnmat,SpinFnature}. There is also clear evidence about the strong spin-lattice coupling \cite{MagnSprb,CaSpinWaveNphys}. Meanwhile, though the electron-phonon ($e$-$p$) coupling alone can not explain the high critical temperature, it is proved that the $e$-$p$ coupling effect definitely plays some role in the SC mechanism \cite{LatticeSCJPCJ,LatticeSCnmat}, probably through the spin-channel \cite{epCnature,epCprb}. Hence a better understanding of the magneto-structural coupling effect and the lattice dynamics in the iron-arsenide parent compounds is important in unraveling the SC pairing mechanism of the iron-based superconductors.

$^{57}$Fe M\"ossbauer spectroscopy (MS) is an excellent probe for both structural and magnetic local properties in iron-containing compounds. In iron-based superconductors the iron element serves as the primary constituent, which makes MS a useful analytical tool \cite{MossPhysicaC}, especially for investigating the lattice dynamics in these systems. Indeed, MS has been widely used to study the iron-based superconductors, e.g. ReFeAsO (Re=rare earth elements) \cite{1111-M,1111-M2}, AFe$_2$As$_2$ (A=Ba, Ca, Eu) system \cite{122-M1,122-M2,122-M3} and LiFeAs \cite{111-M}. However, properties regarding magneto-structural coupling effect and lattice dynamics are hardly discussed in most of these works. In the present work, single crystals of CaFe$_2$As$_2$ were synthesized and studied in detail by MS in the temperature range of 18\,K to 290\,K. It is found that upon cooling through 170\,K, CaFe$_2$As$_2$ undergoes a first-order magneto-structural phase transition and changes from a tetragonal ThCr$_2$Si$_2$-type paramagnetic (PM) phase to an orthorhombic AFM phase \cite{XHChenCa,BasicInfoCa}. Our observations suggest that the lattice dynamics of both the tetragonal and orthorbombic phases are very similar in the high-frequency modes, which could be denoted by the almost same Debye temperatures of the two phases.

\section{\label{sec:Experiment}Experiments}
Single crystals of CaFe$_2$As$_2$ were grown by the FeAs self-flux method described elsewhere \cite{XHChenCa}. The resistivity and hall effect analysis were reported earlier \cite{XHChenCa}. The phase purity of the crystalline sample was confirmed by single crystal X-ray diffraction measurements using a Philips X'pert diffractometer with Cu K$_\alpha$ radiation. DC magnetization measurement was carried out through a commercial (Quantum Design) superconducting quantum interference device (SQUID) magnetometer. The transmission M\"ossbauer spectra were recorded using a conventional constant acceleration spectrometer with a $\gamma$-ray source of 25\,mCi $^{57}$Co in palladium matrix moving at room temperature. The absorber was kept static in a temperature-controllable cryostat. All isomer shifts are quoted relative to $\alpha$-Fe at room temperature. The M\"ossbauer spectra are fitted with \textsc{Mosswinn 3.0i} programme \cite{KlencsarZ}.

\section{\label{sec:Results}Results and Discussion}

\begin{figure}[htp]
\includegraphics[width=8 cm]{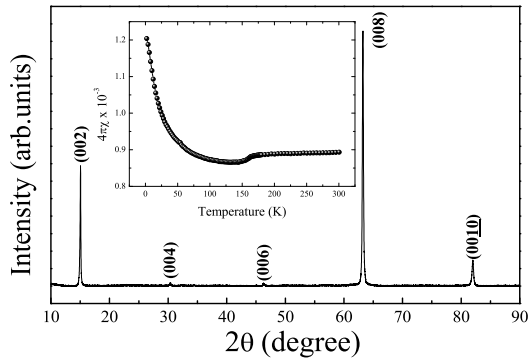}% Here is how to import EPS art
\caption{\label{XRD} Single crystal X-ray diffraction pattern of CaFe$_2$As$_2$. The inset shows the temperature dependence of the magnetic susceptibility measured under H = 5\,T.}
\end{figure}

Fig. \ref{XRD} presents the single crystal X-ray diffraction pattern of the CaFe$_2$As$_2$ compound. The inset shows the temperature dependence of the magnetic susceptibility measured under H = 5\,T. As can be seen, only (00l) diffraction peaks are observed, indicating that the crystallographic c-axis is perpendicular to the plane of the plate-like single crystal. The susceptibility curve shows an anomaly around 165\,K corresponding to the magneto-structural phase transition, which is similar to previous reports \cite{XHChenCa}. All these results indicate that the sample has a good quality and hence a good starting point for the following MS measurements.

\begin{figure}[tbp]
\includegraphics[width=8 cm]{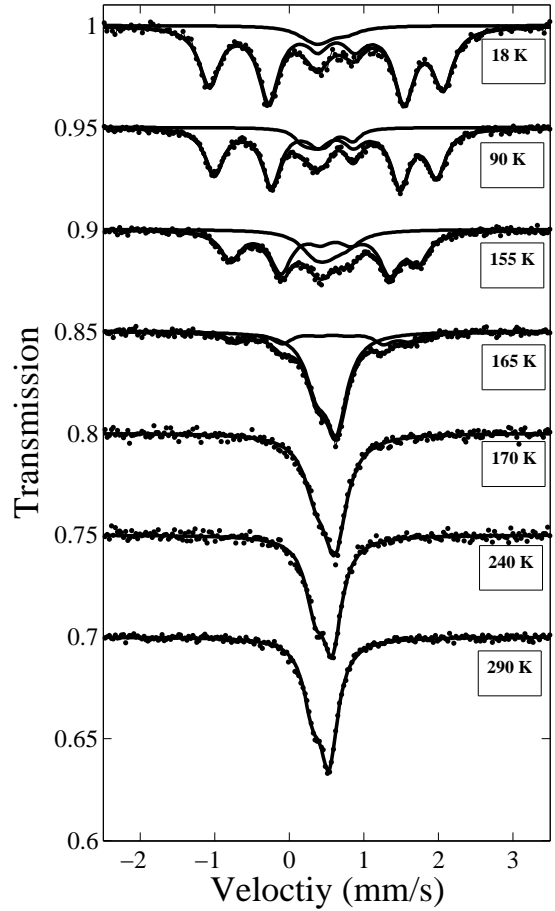}% Here is how to import EPS art
\caption{\label{MS} $^{57}$Fe M\"ossbauer spectra of the single crystal CaFe$_2$As$_2$ obtained at indicated temperatures. The spectra taken between 18\,K-155\,K were fitted with two Zeeman patterns as discussed in the text. The spectra recorded at 160\,K and 165\,K were fitted with a sextet and an asymmetric doublet. The spectra taken above 170\,K were fitted with only one asymmetric doublet.}
\end{figure}

The $^{57}$Fe M\"ossbauer spectra of the single crystal CaFe$_2$As$_2$ recorded in the temperature range of 18\,K-290\,K are shown in Fig. \ref{MS}. Above 170\,K, the spectra are quadrupole split paramagnetic lines and can be well fitted with only one asymmetric doublet. This means that the local environment of the Fe ion is unique, indicating no impurity phases exist in the M\"ossbauer absorber. The spectra recorded at 160\,K and 165\,K were fitted with a sextet corresponding to the AFM phase and an asymmetric doublet corresponding to the residual paramagnetic phase. The coexistence of both phases is characteristic for first-order phase transitions. At temperatures below 155\,K, the spectra can be fitted with the superposition of two Zeeman patterns. Two subspectra are needed to get a better fit to the observed spectra, which is similar to previous reports \cite{Ca-Mossprb}. This may be understood as both commensurate (main component) and incommensurate (minor component) spin density waves (SDW) exist in the sample. The incommensurate SDW has also been observed in Ref \cite{Ca-Mossprb}, where three sextets are used to fit the M\"ossbauer spectrum. Yet the incommensurability in 122-type iron-based parent compounds seems to be controversial. For example, for BaFe$_2$As$_2$ and EuFe$_2$As$_2$ compounds one component is enough to get a good fit to the spectrum and the shape of the SDW is quasirectangular \cite{122-M2} at low temperatures. While the shape of the SDW does not develop a fully quasirectangular shape \cite{122-M2} even at liquid helium temperature, which may be due to the incommensurate SDW. So, understanding the incommensurate SDW certainly deserves much more careful experimental examinations.

\begin{figure}[bhp]
\includegraphics[width=8 cm]{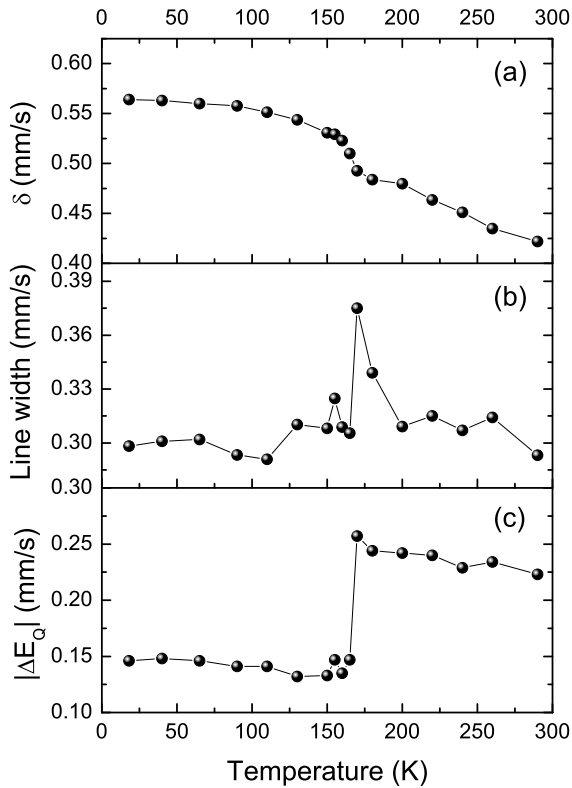}% Here is how to import EPS art
\caption{\label{MSPara} Temperature dependence of the M\"ossbauer hyperfine parameters extracted by fitting the spectra taken at different temperatures. (a) Isomer Shift $\delta$, (b) Line Width $\Gamma$, (c) Quadrupole splitting $\mid\Delta E_Q\mid$.}
\end{figure}

The temperature dependence of hyperfine parameters extracted by fitting the measured M\"ossbauer spectra at different temperatures are plotted in Fig \ref{MSPara}. It is well known that the total isomer shift ($\delta$) is the sum of chemical shift ($\delta_C$) and thermal shift ($\delta_{SOD}$), i.e. $\delta=\delta_C+\delta_{SOD}$. As shown in Fig \ref{MSPara} (a), the isomer shift has a typical value of 0.42\,mm/s for the 122-type iron-based superconductors at room temperature \cite{MossPhysicaC}. Upon cooling, $\delta$ increases gradually before 170\,K, the characteristic of $\delta_{SOD}$, then shows a sudden increase in a narrow temperature range, which could be mainly ascribed to the contribution of $\delta_C$. In M\"ossbauer experiments, $\delta_C$ can be expressed as \cite{ChenMBook} $\delta_C = \alpha(\rho_{\alpha Fe} - \rho_A)$, where $\rho_{\alpha Fe}$ and $\rho_A$ denote the electron density at the iron nucleus for $\alpha$-Fe and the absorber, respectively. $\alpha$ is a calibration constant, which is positive for $^{57}$Fe nucleus. Thus, a positive isomer shift indicates a smaller electron density at the iron nucleus in the sample than in $\alpha$-Fe. Therefore, the sharp increase in $\delta$ at $\sim$165\,K corresponds to a decrease in the electron density at the iron nucleus, which may arises from the changes of the unit volume at the transition temperature. The structural transition results in larger Fe-As distances in the orthorhombic phase, which reduces the interaction between Fe-3d and As-4p electrons and leads to more localized Fe-3d electrons. This will enhance the screening effect of the Fe-3d electrons to s electrons near the nuclear, causing smaller s-electron densities at the iron nucleus thus a bigger isomer shift.

The spectrum line width, $\Gamma$, and quadrupole splitting, $\mid\Delta E_Q\mid$, are shown in Fig \ref{MSPara} (b) and (c), respectively. They both show an anomaly around 165\,K, which is ascribed to the magneto-structural phase transition. One should notice that the line broadening actually appears at temperatures above the phase transition. Similar line broadening effect has been observed in (Ba$_{1-x}$K$_x$)Fe$_2$As$_2$ systems \cite{MRotter2009}. The most probable reason for the line broadening effect can be considered as the onset of magnetic order and/or short range magnetic fluctuations that presents in the CaFe$_2$As$_2$ compound \cite{CaSpinWaveNphys}. These observations highly suppose the magnetic origin of the magneto-structural phase transition in CaFe$_2$As$_2$ parent compound. Upon cooling, some Fe spins happen to form clusters with short-ranged magnetic order in the nonmagnetic matrix by spin fluctuations of the Fe-3d charges. These spin clusters broaden the line width instead of detectable hyperfine field. With further cooling, the phase transition happens when the spin clusters reaches its critical point and causes structural distortion simultaneously to optimize the energy of the system due to spin-lattice coupling effect. Therefore, it is suggested that the magnetic fluctuations and spin-lattice coupling effect may be responsible for the magneto-structural phase transition in CaFe$_2$As$_2$ compound.

\begin{figure}[htp]
\includegraphics[width=8 cm]{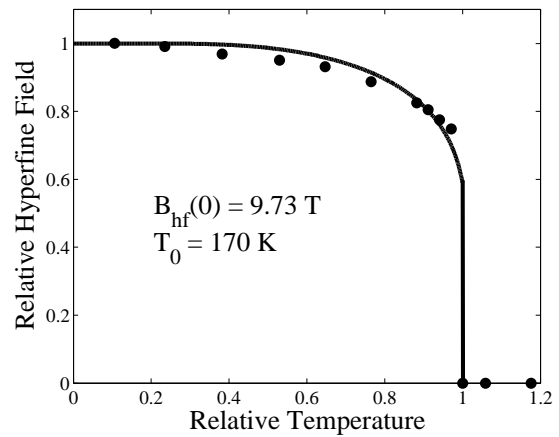}% Here is how to import EPS art
\caption{\label{BhfT} The reduced hyperfine magnetic field vs reduced temperature for the single crystal CaFe$_2$As$_2$ sample. The solid line is theoretical fit to the experimental data obtained on heating, assuming $T_0$=170\,K (see text).}
\end{figure}

To examine the magnetic properties and get insight into the AFM phase transition, the temperature dependence of the hyperfine magnetic field is investigated using a mean field model after Bean and Rodbell \cite{BeanRodbell}, which, while describing the order parameter variation correctly, does not yield information about the magentic fluctuations in side of the hysteretic region, observed here. The reduced hyperfine magnetic field vs reduced temperature is plotted in Fig \ref{BhfT} and the solid line is the calculated theoretical curve. In the Bean-Rodbell model, the temperature dependence of the sublattice magnetization in a AFM material is expressed as \cite{BeanRodbell} $T/T_0 = (\sigma/\tanh^{-1}\sigma)(1+\eta \sigma^2/3)$ in the low spin state approximation. $\eta=\frac{3}{2}Nk_BKT_0\beta^2$, where $\sigma$ is the reduced sublattice magnetization, $\eta$ is a fitting parameter ($\eta<1$ for a second order transition, $\eta>1$ for a first order transition and for $\eta=0$ the equation reduces to a Brillouin function), $\beta$ is the slope of the dependence of transition temperature on volume, $T_0$ would be the transition temperature if the lattice were not compressible, $K$ is the compressibility, $T$ is the temperature and $k_B$ is the Boltzmann constant. As can be seen, Fig \ref{BhfT} shows an approximate agreement between the theory and experimental data. The observed hyperfine field decreases more faster with increasing temperature than the theory predicts in the low temperature range, which could be explained by the low-lying spin excitations as evidenced by inelastic neutron scattering \cite{CaSpinWaveNphys}. The fitted values of $\eta$ and $B_{hf}(0)$ with fixed $T_0$=170\,K are found to be 1.35 and 9.73\,T, respectively. The value of $\eta$ indicates that the nature of the AFM transition is of first order. This is in agreement with previously reported thermodynamic, transport and microscopic data \cite{XHChenCa}. And the fitted zero point hyperfine field is very close to the value ($B_{hf}$$\sim$10\,T) reported by Kumar et al. \cite{Ca-Mossprb} at 4.2\,K.

If we assume the Bean-Rodbell model to be valid, we may determine the compressibility of the CaFe$_2$As$_2$ compound. According to the Bean-Rodbell model, the compressibility is given by $K=6(\Delta v/v_0)^2/(\eta Nk_BT_0\sigma^4)$, where $\Delta v/v_0$ is the volume change at the transition. If we take from experimental values, the volume change $\Delta v/v_0 = -0.93\%$ \cite{VolumeChangeprb} (The negative sign of $\Delta v/v_0$ means a compression of the lattice during the AFM/PM transition, which is consistent with the Bethe-Slater curve.),  $T_0=170\,K$, $\sigma=0.77$ (taken at T=165\,K), $\eta=1.35$ and, assuming $N=2.79\times10^{28}\,m^{-3}$ (corresponding a density of 7\,$g/cm^{3}$), one gets $K=1.67\times10^{-2}\,GPa^{-1}$. The results is very close to the value of first principles calculations on CaFe$_2$As$_2$ ($K=1.65\times10^{-2}\,GPa^{-1}$) \cite{CompCalc}, while a little larger than that of the reported experimental values, $1.04\times10^{-2}\,GPa^{-1}$ for the LaFePO \cite{ComLaFePO} and $0.98\times10^{-2}\,GPa^{-1}$ for the Nd(O$_{0.88}$F$_{0.12}$)FeAs \cite{ComNdFeAsOF}. Thus a direct measurement of the compressibility would be a useful check on this theory.

As discussed above, the magneto-structural phase transition in CaFe$_2$As$_2$ is first-order in nature. In order to obtain a theoretical interpretation of our experiments, we adopt the phenomenological Landau theory of phase transitions to the system under study. In Landau's theory the free-energy density of the system is expanded in terms of two order parameters of $\sigma$ (sublattice magnetization) and $\varepsilon$ (strain) as \cite{LandauFEq1}
\begin{eqnarray}
F(\sigma,\varepsilon) = \frac{1}{2}a(T)\sigma^2 + \frac{1}{4}b\sigma^4 + \frac{1}{6}c\sigma^6
+ \frac{1}{2}B\varepsilon^2 + \lambda\varepsilon\cdot \sigma^2.
\label{Landau1}
\end{eqnarray}
Where $B$ is the elastic modulus and $\lambda$ is the magneto-elastic coupling coefficient. Minimizing the total energy with respect to the strain yields $\varepsilon=-\lambda \sigma^2/B$. Substituting $\varepsilon$ with $\sigma$ in Eq. (\ref{Landau1}), one obtains a renormalized 2-4-6 Landau free energy,
\begin{eqnarray}
F(\sigma) = \frac{1}{2}a(T)\sigma^2 + (\frac{1}{4}b-\frac{\lambda^2}{2B})\sigma^4 + \frac{1}{6}c\sigma^6.
\label{Landau2}
\end{eqnarray}
The magneto-elastic coupling is presented by the fourth order term. To estimate the coefficient of this term, we turn to the Bean-Rodbell model, from which the magnetic part of the free energy without external field is also a 2-4-6 function of $\sigma$, as \cite{BeanJAPEq3}
\begin{eqnarray}
G(\sigma) = \frac{1}{2}(T-T_0)\sigma^2 + \frac{1}{12}(T-\eta T_0)\sigma^4 + \frac{1}{6}c\sigma^6.
\label{BeanG}
\end{eqnarray}
Comparing Eq. (\ref{Landau2}) and (\ref{BeanG}), clearly, the sign of $(b/4 - \lambda^2/2B)$ is determined by $(T-\eta T_0)$, which is negative when $\eta=1.35$ around the transition temperature. A negative fourth order term in Landau free energy creates an energy barrier in the free energy landscape, this leads to a first-order transition. Obviously, a large magneto-elastic coupling coefficient $\lambda$ results in a large negative fourth order term. The more negative is this term, the larger is the transition barrier, and the more distinct are the first-order transition characters. On the other hand, a larger $\lambda$ leads to a larger spontaneous lattice distortion upon the magnetic transition. Therefore, the strength of the magneto-elastic coupling $\lambda$ can be represented by the magnitude of the spontaneous lattice distortion, like $\eta$, which involves the parameters $K$ and $\beta$ that are related to the volume change. $\eta$ also controls the order of the magnetic phase transition. According to Eq. (\ref{BeanG}), one always gets a negative fourth order term when $\eta > 1$, indicating a first-order magnetic phase transition at $T_0$ due to the energy barrier from magneto-elastic coupling. Thus, the parameter $\eta$ predicts the strength of the coupling.

To investigate the lattice dynamics of the Fe atoms, the temperature dependence of the Lamb-M\"ossbauer factor or recoil-free fraction, $f$, is discussed in what follows. In Debye model of the lattice dynamics, $f$ can be expressed as \cite{ChenMBook}
\begin{eqnarray}
f(T) = e^{\frac{-6E_R}{k_B\Theta_D}[\frac{1}{4} + (\frac{T}{\Theta_D})^2 \int_0^{\Theta_D/T}\frac{x dx}{(e^x - 1)} ]},
\label{LMFactor}
\end{eqnarray}
where $k_B$ is Boltzmann's constant, $\Theta_D$ is the Debye temperature, and $E_R$ is the recoil energy of an free emitting nucleus. In high temperature ($T>\Theta_D/2$) approximation, we may rewrite Eq. (\ref{LMFactor}) as the following formula \cite{ChenMBook}:
\begin{eqnarray}
-ln(f(T)) = \frac{6E_RT}{k_B\Theta_D^2}(1+\epsilon T+\delta T^2+\cdots).
\label{Shortf}
\end{eqnarray}
Here $\epsilon$ and $\delta$ are termed as the anharmonic coefficient and they are both very small values.

\begin{figure}[htp]
\includegraphics[width=8 cm]{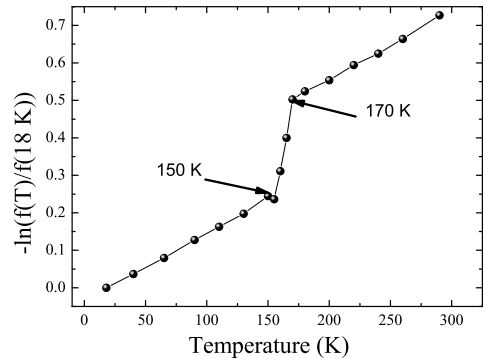}% Here is how to import EPS art
\caption{\label{Area} Logarithm of the relative $f$ factor versus temperature.}
\end{figure}

The temperature dependence of $-ln(f(T)/f(18\,K))$ is shown in Fig \ref{Area}. It is evident that upon cooling, the curve drops steeply through 170\,K to 150\,K due to the occurrence of the phase transition. This sudden increase in $f(T)$ is likely due to the thickness effect, i.e. the increase of spectral area when overlapping peaks separate from each other, as it usually happens when a magnetic phase transition takes place \cite{KlencsarZ}. The temperature dependence of $f(T)$ beyond this temperature region is quite linear, i.e. Fe atoms experience harmonic vibrations in both tetragonal phase and orthorhombic phase. Since the anharmonic effect characterized by $\epsilon$ and $\delta$ is negligible, the $f(T)$ data can be simply fitted with $-ln(f(T)/f(18\,K))=6E_RT/k_B\Theta_D^2 + ln(f(18\,K))$, where $ln(f(18\,K))$ was treated as a free parameter. The best-fit of the experimental data shown in Fig. \ref{Area} yields the following results: $\Theta_D = 272\,K$ for the segment below the phase transition and $\Theta_D = 271\,K$ for the segment above the phase transition. The nearly same behaviors of $f(T)$ with temperature indicates that no obvious changes of the phonon modes in high frequency ranges of the Fe vibrational state, which is in agreement with inelastic neutron scattering results \cite{PhononCaprb} on the phonon spectrum. Moreover, the observed harmonic vibrational behavior of the iron atoms indicates that the large phonon linewidths in CaFe$_2$As$_2$ observed by neutron scattering \cite{CaPhonon} may be due to to strong electron-phonon coupling effect. Furthermore, theoretical calculations of the phonon spectrum show that including magnetism does not improve agreement with experiments \cite{CaPhonon} for CaFe$_2$As$_2$. This is rather different for other 122-type, SrFeAsF \cite{newphonon} and LaFeAsO \cite{LaFeAsOphonon} 1111-type systems, where the authors show that including magnetic degree of freedom greatly improves the agreements with experiments. Thus more work is necessary to understand these properties.

\section{\label{sec:Conclusion}Concluding remarks}
To conclude, we have studied the structural and magnetic properties of CaFe$_2$As$_2$ single crystal in detail by M\"ossbauer spectroscopy. Our results show that the structural distortion and magnetic transition are coalesced into one single first-order magneto-structural phase transition through strong magneto-structural coupling effect. Short-range magnetic order and/or spin fluctuations appears well above the magneto-structural phase transition temperature and may be responsible for the phase transition. Analyzing the temperature dependence of the hyperfine magnetic field in the Bean-Rodbell model, we are able to determine the compressibility of the CaFe$_2$As$_2$ crystal, $K = 1.67\times10^{-2}\,GPa^{-1}$.  Both the PM and AFM phases exhibit similar lattice dynamic behaviors of the Fe atoms in high-frequency modes and they both have a Debye temperature close to $\sim$270\,K.

\begin{acknowledgments}
The authors are grateful to X.H. Chen and his group for supplying the sample. One of the author would also like to thank Z. Klencs\'{a}r for useful discussions. This work is supported by the National Natural Science Foundation of China (Grant No: 10975066).

\end{acknowledgments}

%\begin{thebibliography}{00}

%\end{thebibliography}

%\bibliography{MyRef}

%%%%%%%%%%%%%%%%%%%%%%%%%%%%%%%%%
%Merlin.mbs v4.21 2009-07-09.
%\section*{References}

%\bibliography{MyRef}

\begin{thebibliography}{00}
\bibitem{26KLaFeAsOF} Kamihara Y, Watanabe T, Hirano M and Hosono H 2008 J. Am. Chem. Soc. 130 3296
\bibitem{ReviewNature} Paglione J and Greene R L 2010 Nature Phys. 6 645
\bibitem{ReviewMagn} Lumsden M D and Christianson A D 2010 J. Phys.: Condens. Matter 22 203203
\bibitem{ThePuzzles} Johnston D C 2010 Adv. Phys. 59 803
\bibitem{SpinFnmat} Drew A J, Niedermayer C, Baker P J, Pratt F L, Blundell S J, Lancaster T, Liu R H, Wu G, Chen X H, Watanabe I, Malik V K, Dubroka A, R\"ossle M, Kim K W, Baines C and Bernhard C 2009 Nature Mater. 3 310

\bibitem{SpinFnature} Cruz C, Huang Q, Lynn J W, Li J, Ratcliff W, Zarestky J L, Mook H A, Chen G F, Luo J L, Wang N L and Dai P 2008 Nature 453 899
\bibitem{MagnSprb} Jesche A, Canales N C, Rosner H, Borrmann H, Ormeci A, Kasinathan D, Klauss H H, Luetkens H, Khasanov R, Amato A, Hoser A, Kaneko K, Krellner C and Geibel C 2008 Phys. Rev. B 78 180504(R)
\bibitem{CaSpinWaveNphys} Zhao J, Adroja T, Yao D X, Bewley R, Li S, Wang X F, Wu G, Chen X H, Hu J and Dai P 2009 Nature Phys. 5 555
\bibitem{LatticeSCJPCJ} Lee C, Iyo A, Eisaki H, Kito H, Fernandez-Diaz M T, Ito T, Kihou K, Matsuhata H, Braden M and Yamada K 2008 J. Phys. Soc. Japn. 77 083704
\bibitem{LatticeSCnmat} Zhao J, Huang Q, Cruz C L, Li S, Lynn J W, Chen Y, Green M A, Chen G F, Li G, Li Z, Luo J L, Wang N L and Dai P 2008 Nature Phys. 7 953

\bibitem{epCnature} Liu R H, Wu T, Wu G, Chen H, Wang X F, Xie Y L, Ying J J, Yan Y J, Li Q J, Shi B C, Chu W S, Wu Z Y and Chen X H 2009 Nature 459 64
\bibitem{epCprb} Boeri L, Calandra M, Mazin I I, Dolgov O V and Mauri F 2010 Phys. Rev. B 82 020506
\bibitem{MossPhysicaC} Nowik I and Felner I 2009 Physica C 469 485

\bibitem{1111-M} Kitao S, Kobayashi Y, Higashitaniguchi S, Saito M, Kamihara Y, Hirano M, Mitsui T, Hosono H and Seto M 2008 J. Phys. Soc. Jpn. 77 103706
\bibitem{1111-M2} McGuire M A, Hermann R P, Sefat A S, Sales B C, Jin R, Mandrus D, Grandiean F and Long G J 2009 New J. Phys. 11 025011
\bibitem{122-M1} Rotter M, Tegel M, Schellenberg I, Schappacher F M, P\"ottgen R, Deisenhofer J, G\"unther A, Schrettle F, Loidl A and Johrendt D 2009 New J. Phys. 11 025014
\bibitem{122-M2} Blachowski A, Ruebenbauer K, Zukrowski J, Rogacki K, Bukowski Z and Karpinski J 2011 Phys. Rev. B 83 134410
\bibitem{122-M3} Rotter M, Tegel M, Johrendt D, Schellenberg I, Hermes W and P\"ottgen R 2008 Phys. Rev. B 78 020503
\bibitem{111-M} Gao W B, Linden J, Wang X C, Jin C Q, Tohyama T, Karppinen M and Yamauchi H 2010 Solid State Communications 150 1525

\bibitem{XHChenCa} Wu G, Chen H, Wu T, Xie Y L, Yan Y J, Liu H R, Wang X F, Ying J J and Chen X H 2008 J. Phys.: Condens. Matter 20 42220
\bibitem{BasicInfoCa} Ni N, Nandi S, Kreyssig A, Goldman A I, Mun E D, Bud'ko S L and Canfield P C 2008 Phys. Rev. B 78 014523


\bibitem{KlencsarZ} Klencs\'{a}r Z, Kuzmann E, Homonnay Z, W\'{e}rtes A, Simopoulos A, Devlin E and Kallias G 2003 J. Phys. Chem. Solids 64 325
\bibitem{Ca-Mossprb} Kumar N, Nagalakshmi R, Kulkarni R, Paulose P L, Nigam A K, Dhar S K and Thamizhavel A 2009 Phys. Rev. B 79 012504


\bibitem{ChenMBook} Chen Y L and Yang D P 2007 M\"ossbauer Effect in Lattie Dynamics: Experimental Techniques and Applications (Wiley-VCH Verlag)
\bibitem{MRotter2009} Rotter M, Tegel M, Schellenberg I, Schappacher F M, P\"ottgen R, Deisenhofer J, G\"unther A, Schrettle F, Loidl A and Johrendt D 2009 New J. Phys. 11 025014
\bibitem{BeanRodbell} Bean C P and Rodbell D S 1962 Phys. Rev. 126 104

\bibitem{VolumeChangeprb} Goldman A I, Argyriou D N, Ouladdiaf B, Chatterji T, Kreyssig A, Nandi S, Ni N, Dud'ko S L, Canfield P C and McQueeney R J 2008 Phys. Rev. B 78 100506(R)
\bibitem{CompCalc} Wang Y, Ding Y and Ni J 2009 Solid State Commun. 149 2125
\bibitem{ComLaFePO} Igawa K, Okada H, Arii K, Takahashi H, Kamihara Y, Hirano M, Hosono H, Nakano S and Kikegawa T 2008 J. Phys. Soc. Jpn. 78 023701
\bibitem{ComNdFeAsOF} Zhao J, Wang L, Dong D, Liu Z, Liu H, Chen G, Wu D, Luo J, Wang N, Yu Y, Jin C and Guo Q 2008 J. Am. Chem. Soc. 130 13828


\bibitem{LandauFEq1} Yang S and Ren X 2008 Phys. Rev. B 77 014407
\bibitem{BeanJAPEq3} Zach R, Guillot M and Tobola J 1998 J. Appl. Phys. 83 7237
\bibitem{PhononCaprb} Mittal R, Rols S, Zbiri M, Su Y, Schober H, Chaplot S L, Johnson M, Tegel M, Chatterji T, Matsuishi S, Hosono H, Johrendt D and Brueckel T 2009 Phys. Rev. B 79 144516

\bibitem{CaPhonon} Mittal R, Pintschovius L, Lamago D, Heid R, Bohnen K P, Reznik D, Chaplot L, Su Y, Kumar N, Dhar S K, Thamizhavel A and Brueckel T 2009   Phys. Rev. Lett. 102 217001
\bibitem{newphonon} Zbiri M, Mittal R, Rols S, Su Y, Xiao Y, Schober H, Chaplot S L, Johnson M R, Chatterji T, Inoue Y, Matsuishi S, Hosono H and Brueckel T 2011 J. Phys.: Condens. Matter 22 315701

\bibitem{LaFeAsOphonon} Higashitaniguchi s, Seto M, Kitao S, Kobayashi Y, Saito M, Mauda R, Mitsui T, Yoda Y, Kamihara Y, Hirano M and Hosono H 2008 Phys. Rev. B 77 174507

\end{thebibliography}

\end{document}